# Harmonic-Induced Plasmonic Resonant Energy Transfer between Metal and Semiconductor Nanoparticles

Yueming Yan, Nathan J. Spear, Adam J. Cummings, Karina Khusainova, Janet E. Macdonald, Richard F. Haglund*

Y. Yan, A. J. Cummings, K. Khusainova, R. F. Haglund
Department of Physics and Astronomy
Vanderbilt University
Nashville, Tennessee 37235 United States of America
E-mail: richard.haglund@vanderbilt.edu

N. J. Spear
Vanderbilt Institute of Nanoscale Science and Engineering
Vanderbilt University
Nashville, Tennessee 37235, United States of America

J. E. Macdonald,
Department of Chemistry
Vanderbilt Institute of Nanoscale Science and Engineering
Vanderbilt University
Nashville, Tennessee 37235 United States of America

ORCID
Yueming Yan: 0000-0002-2158-0177
Nathan J. Spear: 0000-0002-0881-6798
Adam Cummings: 0009-0004-6672-6944
Karina Khusainova: 0009-0005-9767-8508
Janet E. Macdonald: 0000-0001-6256-0706
Richard F. Haglund: 0000-0002-2701-1768

**Abstract**

Heterostructures combining two or more metal and/or semiconductor nanoparticles exhibit enhanced upconversion arising from localized surface plasmon resonances (LSPRs). However, coupled plasmon-exciton systems are slowed by excitonic relaxation and metallic multi-plasmon systems are not broadly tunable. Here, we describe a heterostructure in which insulating alumina layers vary separation between CuS and Au nanoparticles, allowing experimental confirmation of the $d^{-6}$ dependence typical of surface-dipole mediated interactions between Au and CuS plasmons, as demonstrated in Lumerical® simulations. Transient-absorption spectroscopy shows faster plasmon relaxation in heterostructured Au/CuS (690 fs) than CuS nanoparticles (929 fs), signifying direct energy transfer. Moreover, coupling between the second-harmonic LSPRs of CuS and the fundamental LSPR in Au is evident in nonlinear absorption measurement. This defines a novel harmonic-induced plasmonic resonant energy transfer (HIPRET) dynamic linking the metallic Au plasmon and the broad semiconductor plasmon in CuS. This prototype for tunable, ultrafast plasmonic upconversion exemplifies a strategy for high-efficiency nonlinear nanodevices that have promising applications in photocatalysis, parametric down-conversion and biomedical imaging.

**Teaser**

This work advances the understanding of near-field ultrafast plasmonic interactions that describes the sub-picosecond upconversion dynamics.

# Introduction

Ultrafast nonlinear optical devices that can upconvert photon frequencies have applications in fields ranging from optical communications to medical diagonostics (1-3). Metal–nanoparticle systems have long been known to exhibit strongly enhanced upconversion effects, such as harmonic generation (4-6) and multiphoton photoluminescence (MPPL) (7-10) mediated by large local electric field generated by localized surface plasmon resonances (LSPRs), collective oscillations of surface charge carriers excited at the resonant plasmon frequency. To further increase nonlinear performance, multi-plasmonic structures featuring harmonically coupled plasmonic resonances — one at the excitation wavelength, the other at a harmonic frequency of that wavelength — have also attracted substantial interest. In such structures, multiple plasmon resonances can be excited by plasmon-plasmon interactions driven at a fundamental resonance frequency, leading to substantial electric-field enhancements at both fundamental and harmonic frequencies. The majority of relevant studies on multi-plasmonic structures are based on metallic nanoparticles (NPs) prepared by lithographic techniques, in which either multipole (11) or multi-modal (12) plasmon modes are excited in a single metallic crystal or by incorporating structures with two different metallic particles in which the plasmon resonance in both metallic components satisfies the harmonic condition (13). However, in metal dual-plasmonic nanostructures, significant challenges include the high cost of lithographic nanofabrication and the inherently narrow bandwidth of metallic nanoparticle LSPRs, which severely limits the tuning range of absorption and emission present.

The broad LSPRs of semiconductor nanoparticles, on the other hand, enable tuning from the visible to the mid-infrared by manipulating size, geometry, and doping level (14,15). A new class of the plasmonic hetero-nanocomposites integrates metal and semiconductor NPs synthesized by solvothermal processes exhibits extraordinary tunability; moreover, such processes are also compatible with to large area preparations by dipcoating or painting, for example. Previous studies show that these metal-semiconductor heterostructures feature strongly enhanced light absorption and are attractive for photothermal therapy and solar energy applications (16-19). An extreme example is Janus metal-semiconductor nanoparticles suspended in liquid, which exhibit ultrafast light absorption due to plasmon-induced hot-electron injection (20,21), but not light emission. However, the liquid environment limits the potential for incorporating such nanostructures into nonlinear frequency conversion devices. Many other studies have reported enhanced upconversion in coupled metal-semiconductor nanoparticle systems by plasmon-exciton coupling (22-24), in which upconverted emission from the semiconductor excitons is boosted by local electric field from proximate metal plasmons. However, the time scales dictated by exciton relaxation are substantially slower than observed in metallic plasmon-plasmon coupling.

Unlike the numerous studies on enhanced light *absorption* in metal-semiconductor plasmonic nanocomposites, we focus here on ultrafast light *emission* following the excitation of semiconductor and metal LSPRs. In Au/CuS dual-plasmonic nanoparticle heterostructure films, spectral overlap between the second harmonic of the CuS LSPR (1050 nm) and the fundamental mode of the Au LSPR (525 nm) allows these dipolar plasmon resonances to be simultaneously excited by two-photon absorption of intense NIR laser pulses (1050nm), coupling and transferring the second harmonic energy to each other *via* surface dipole-dipole interactions that depend on the inverse sixth-power of their separation (25). This plasmonic coupling enables bidirectional resonant energy transfer, leading to enhanced second-harmonic generation (SHG) and third-harmonic generation (THG). We call this process harmonic-induced plasmonic resonant energy transfer (HIPRET) between the plasmonic semiconductor and metal plasmons, to distinguish it from the directional plasmon-induced resonant energy transfer process, in which the plasmonic dipole of metallic NPs excites an electron-hole pair in a proximate semiconductor (26-28).

The HIPRET process occurs on the sub-picosecond time scale at excitation energies below the threshold for exciton creation in the semiconductor and serves as the intermediate stage to enhance even high-order harmonic generation (29). To characterize the mechanism of HIPRET, we deposit alumina films of varying thickness between Au and CuS nanoparticle layers to measure the distance dependence of the harmonic emission. We use finite-difference time-domain (FDTD) simulations in Lumerical® to model the near-field enhancement induced by plasmonic interactions in the Au/CuS heterostructure under 1050 nm excitation. The simulated coupling efficiency matches well to the experimental results, both featuring $d^{-6}$ dependence, and strongly implies that the plasmon-plasmon coupling in the Au/CuS heterostructure is mediated by the surface dipoles. Moreover, ultrafast transient absorption spectroscopy of CuS plasmon dynamics shows that resonant energy transfer between CuS and Au reduces the plasmon relaxation time by about one-third compared to the relaxation time of the CuS plasmon alone. Finally, when the pump frequency is significantly detuned from the fundamental plasmon resonance of CuS, the dominant upconversion mechanism is found to change from harmonic generation to MPPL. This observation establishes the critical importance of the harmonic condition between the LSPRs and the pump wavelength on the efficiency of HIPRET—and thus, the total upconversion signal—in the Au/CuS hybrid films.

**Distance Dependence of Harmonic Intensity**

To explore the mechanism underlying the plasmonic interactions, an approach to accurately control the separation between the CuS and Au layers is desirable. The previously employed bath method of nanoparticle deposition cannot provide precise control over the amount and length of the coordinating ligands between the Au and CuS NPs; thus there was no reliable way to adjust separation at the nanometer scale. Attempts to change the separation distance between the nanoparticle layers by changing the length of the interlayer ligand failed, likely due to the presence of multiple intercalated layers of ligands instead of a single monolayer.

Here, an alternate approach using a spacer layer of alumina was developed. First, the Au NPs are deposited on the microscope slide by spin coating, and then the alumina layer is deposited on top of the Au NPs by electron-beam evaporation. The high temperature and vacuum conditions during alumina deposition facilitates the removal of ligands attached to the nanoparticles. Finally, the CuS NPs are deposited atop the alumina layer by spin coating (Fig. 1a). Alumina films of 5 nm, 6 nm, 10 nm, 30 nm, 50 nm, 60 nm thicknesses are confirmed by atomic force microscopy (AFM). The uniformity of the $Al_2O_3$ layer deposition on Au films is characterized by energy dispersive spectroscopy (EDS) and the nearly uniform distribution of the $Al_2O_3$ layers makes it reasonable to use the thickness of the $Al_2O_3$ layer to approximate the separation distance between Au and CuS films (Figs. S1a-c). The absorption spectrum of the Au-$Al_2O_3$-CuS structure is measured by UV-vis-NIR spectrophotometry (Fig. 1b). The existence of the $Al_2O_3$ layers does not shift the dual-LSPR features in the hybrid structure.

To experimentally explore the distance-dependence of harmonic generation $Al_2O_3$ spacer layers of varying thickness were used. The harmonic generation measurement scheme for the Au/CuS films has been described previously (30,31). Each Au/CuS bilayer nanoparticle film with its defined separation between the Au and CuS layers is exposed to 150 fs pulses from a mode-locked Nd:glass laser operating at 1050 nm. The plots of the SHG and THG intensities at the peak excitation laser intensity are calculated as a function of spacer thickness (Figs. 1c and 1d). Both the SHG and THG peak intensities decrease dramatically as the separation distance increases from 5 nm to 10 nm and reveal a sharp inflection for distances larger than 10 nm. Beyond this distance, the SHG and THG intensities do not vary significantly, and the bilayer films produce a relatively low overall harmonic generation signal. We interpret these results to indicate that at separations beyond 10 nm, Au and CuS NPs act independently, and the intense local electric fields from their surface plasmon resonances are effectively shielded from each other by the insulating alumina layers so that the plasmon-plasmon coupling is greatly depressed. At smaller distances, the local electric fields from the surface plasmons spatially overlap and interact. This demonstrated effective distance dependence is consistent with the classic Förster or fluorescence resonant energy transfer (FRET, 1-10 nm) (32), suggesting a dipolar resonant energy transfer mechanism.

We find that the SHG peak intensity *vs* distance is indeed best fit by the function $I_{SHG} = 0.0006 + \frac{57}{d^6}$, and the THG peak intensity *vs* distance by $I_{THG} = 0.0003 + \frac{53}{d^6}$. The fitting parameters 0.0006 and 0.0003 in the functions correspond, respectively, to the incoherent sum of SHG and THG intensities from Au and CuS nanoparticles without plasmonic coupling, consistent with the measurements at large spacer thickness (Figs. 1c and 1d) and equations (1) and (2) below. The fitting parameters 57 and 53 represent the plasmonic coupling strength in the respective plasmon-enhanced harmonic generation, accounting for various coupling factors including the transition dipoles and orientation factors (Sec S2) (33). The comparable coupling strengths suggest that the plasmonic interactions shape both the SHG and THG output through a similar resonant energy transfer process. The strong agreement ($R^2$) between experimental results and the fits indicates that a $d^{-6}$ dependence as well as our theoretical framework is a good approximation of the plasmonic coupling between the Au and CuS NPs.

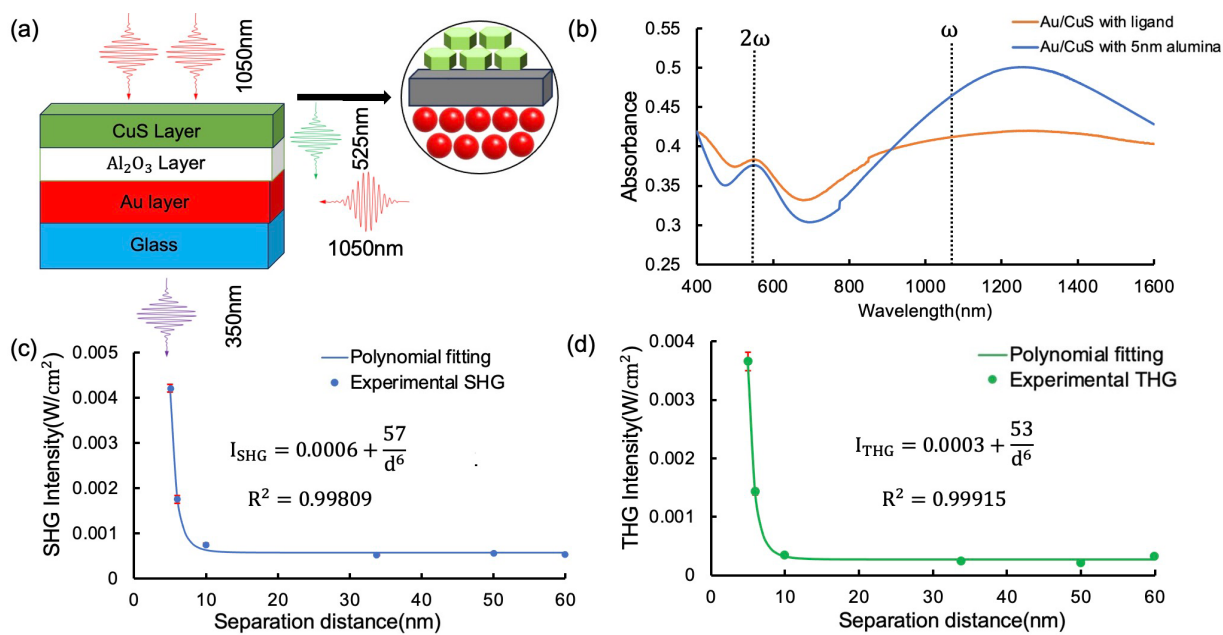

Figure 1. (a) Schematic diagram of the Au/Al$_2$O$_3$/CuS structure illustrating the cascaded THG process enhanced by the coupling from CuS to Au nanoparticles. (b) UV-vis-NIR spectrophotometry of the heterostructure films with different interstitial layers: Au/CuS with 5 nm alumina (blue), Au/CuS with ligand (orange). The peak harmonic generation intensity

with error bas included as a function of the separation distance along with the polynomial fitting function. (c) SHG and (d) THG at pump intensity 0.07 GW/cm$^2$.

The dependence of harmonic generation on the distance between the nanoparticle layers can be calculated using a model of plasmon coupling dynamics. In Section S2, we develop theoretical expressions for the enhanced THG and SHG mediated by plasmonic interactions. When the CuS and Au NPs are irradiated at high laser intensity, the second-harmonic polarization of the CuS plasmon resonance dipoles and the Au plasmon resonance dipoles are induced via two-photon absorption, contributing to the production of second-harmonic light. These dipolar resonances can couple and transfer second-harmonic energy through non-radiative near-field interactions. This interchange of second-harmonic energy from nearby plasmons can either contribute to the emission of second-harmonic light or combine with a photon from the pump laser to generate third-harmonic light by cascaded THG: $2\omega + \omega = 3\omega$. The SHG and THG intensities from Au/CuS heterostructure films can be expressed as a function of separation distance between the CuS and Au NPs as follows:

$$I_{SHG}^{Au/CuS}(2\omega) = [I_{pump}^{Au}(2\omega) + I_{pump}^{CuS}(2\omega)] + \frac{A_{CuS-Au} + A_{Au-CuS}}{d^6} \quad (1)$$

$$I_{THG}^{Au/CuS}(3\omega) = [I_{pump}^{Au}(3\omega) + I_{pump}^{CuS}(3\omega)] + \frac{(A_{CuS-Au}B_{Au} + A_{Au-CuS}B_{CuS})I(\omega)}{d^6} \quad (2)$$

where $I(\omega)$ is the pump intensity, $A_{CuS-Au}$ ($A_{CuS-Au}$) is the coefficient of coupling from CuS (Au) to Au (CuS), and $B_{Au}$ ($B_{CuS}$) is a combined factor relating the efficiency of sum-frequency generation in Au (CuS). In our theoretical model, both $I_{SHG}^{Au/CuS}$ and $I_{THG}^{Au/CuS}$ consist of two terms: the first term is the incoherent sum of harmonic emission from Au and CuS nanoparticles independently, without the coupling effect; the second term represents the enhanced harmonic signal produced by the plasmonic interaction, which displays an inverse-sixth power distance dependence ($d^{-6}$), consistent with a dipole-dipole interactions. Both the second- and third-harmonic light from the experiments and theory exhibit inverse-sixth power dependence, strongly indicating that the harmonic generation enhancement effect observed in the dual-plasmonic heterostructure arises from plasmon-plasmon interactions.

**Plasmonic Near-field Enhancement**

We carried out finite-difference time-domain (FDTD) simulations in Lumerical® to further investigate the local electric field enhancement due to coupling of the Au and CuS surface plasmons. To model the harmonic plasmonic resonance, the polarization of the incident light is set parallel to the basal plane of CuS, in which the NIR in-plane mode of CuS LSPRs dominates (Fig. S3a). The simulated extinction spectrum coincides with experimental results, showing distinct LSPRs at wavelengths of 1050 nm and 525 nm from CuS and Au, respectively (Fig. S3b). In our simulation, the CuS is aligned such that its basal plane is parallel to the Au surface, which reproduces the physical orientation (31). The two nanoparticles are moved along the diagonal with increasing separation distance, so that the orientation of the CuS edge with respect to the Au plasmon field is unchanged (Fig. 2a). The simulation also shows that an offset along the y-axis is needed between the Au and CuS to facilitate direct interference of their optical near fields. When the Au and CuS particles are collinear with the excitation direction (Fig. S4), the enhanced fields at the vertices of the particles do not overlap and the near-field distribution demonstrates extremely weak coupling at the smallest separation distances (2 nm), after which there is no field enhancement between the two NPs. When there is an offset between the CuS and Au NPs, there is substantial spatial overlap of the near fields.

Under 1050 nm excitation, we compute the electric field distribution for 2 nm, 3 nm, 4 nm, 6 nm, 10 nm, 15 nm and 20 nm separation distance—as defined from the edge of the CuS to the nearest surface of the Au sphere (Fig. 2c). As derived in section S2, the harmonic-induced resonant energy transfer rate is proportional to the square of the near-field strength induced by coupled plasmon resonances:

$$W_{\text{HIPRET}}(\omega) = \frac{\eta_n}{\tau_n} \cdot \frac{9c^4}{8\pi} \cdot \left|\frac{\overrightarrow{E_{nf}}}{\mu_n}\right|^2 \cdot \int d\omega \frac{\epsilon_{\text{Au}}(\omega) F_{\text{CuS}}(\omega)}{(\omega)^4} \tag{3}$$

where $n$=Au, CuS, respectively, for the bidirectional energy transfer process, $\eta_n$ is the conversion efficiency of second-harmonic generation of donor nanoparticles, $\tau_n$ is the donor plasmon lifetime, $\mu_n$ is the transition dipole, and $F_{\text{CuS}}(\omega)$ and $\epsilon_{\text{Au}}(\omega)$ represent the plasmon resonance spectra of CuS and Au, respectively. Thus, the distance dependence of the square of the near-field enhancement (SNFE), directly obtained from the simulation could provide a critical clue to the nature of the plasmon-plasmon coupling. The near-field enhancement is measured at a point 1 nm from the Au surface on the diagonal between the nanoparticles and is fixed for all the separation cases. The measurement position was chosen within the effective plasmonic coupling distance (10 nm) closer to the Au, between the Au and CuS NPs, to avoid any influences from hot spots at the corners of the CuS NP (Figure 2c). The best polynomial fit for the distance-dependent SNFE plot: $\left(\frac{E_{\text{nf}}}{E_0}\right)^2 = 4.02 + \frac{4110}{d^6}$ gives an inverse-sixth power distance dependence matching that of the experimental finding for SHG and THG measurements. The fitting parameter 4.02 in the function indicates the Au NPs near-field enhancement effect when the separation distance is sufficiently large that the two NPs function as the independent dipoles. When the distance separating CuS and Au NPs is set to larger than 6 nm in the calculation, only weak near-field enhancement is observed.

Additionally, we conducted simulations where the separation distance between the Au and CuS was changed by moving the nanoparticle along the *x*- (Fig. S5) and *y*- axes independently (Fig. S6). The distance dependence of the calculated SNFE also fits an inverse-sixth power distance dependence (Figs. S7). The agreement demonstrated in these results regardless of different relative orientation strongly suggests the direct coupling of the surface dipoles between the Au and CuS NPs as a general mechanism for the plasmonic interactions.

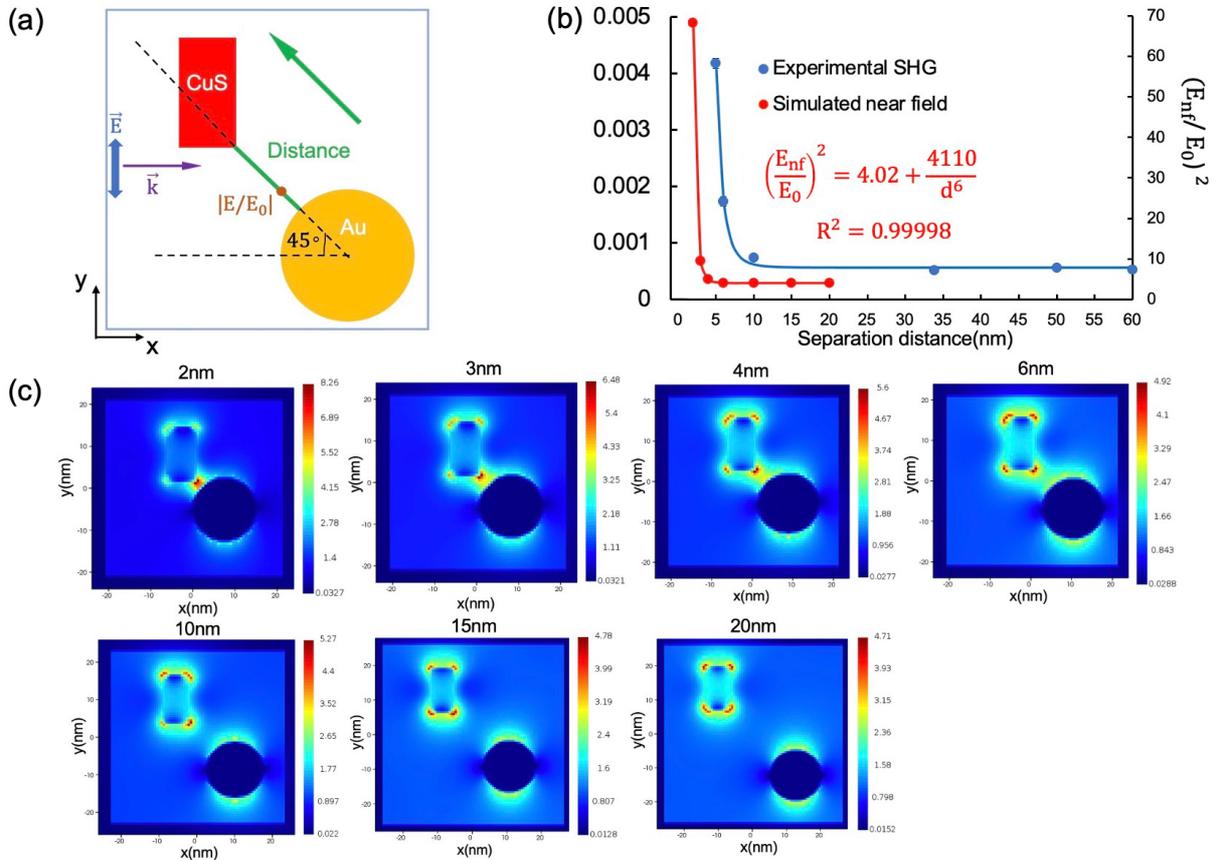

**Figure 2**. (a) Structural model of Au/CuS nanoparticles in the FDTD simulation. (b) The experimental SHG intensity and simulated SNFE as a function of the separation distance with the inverse-sixth power polynomial fit. (c) FDTD calculated local electric field distribution with the varying separation distance between the Au and CuS nanoparticles. The color scale bars show the relative increase in the electric field $|E/E_0|$.

In Fig. 2b, we also present a comparison between the experimental and simulated results. Both agree on the distance dependence as stated previously. However, there is a discrepancy between the inflection distances: 6 nm for the simulation and 10 nm for the experiments. The discrepancy can be attributed to the non-uniform deposition of the alumina layers on the Au films, as observed in the SEM images (Figs. S1d and e). Although EDS confirms that most of the Au NPs are covered by the deposited alumina layers (Figs. S1b and S1c), some small Au NPs (the bright spots) remain uncovered by the alumina and are thus directly in contact with CuS. As a result, the effective average distance between the Au and CuS NPs in our structure is smaller than the alumina layer thickness (Fig. 2b). Despite this limitation, the inverse sixth-power distance dependence of plasmon-plasmon coupling efficiency, seen in both experimental far-field harmonic radiation and theoretical near-field enhancement, identifies plasmonic interaction mechanism as dipole-dipole coupling.

**Ultrafast Plasmon Dynamics**

To further investigate the resonant energy transfer between CuS and Au plasmons, the relaxation dynamics in Au/CuS and CuS are investigated by ultrafast pump-probe measurements (Fig. S8). The absorption spectrum of Au/CuS includes a broad peak spanning wavelengths from 900 nm to 2600 nm that corresponds to the CuS plasmon (Fig. S10). The films are pumped at 1200 nm and probed at 2400 nm by synchronized beams from a Light Conversion Topas optical parametric amplifier (OPA) with a common amplified Ti:sapphire pump to obtain the dynamics of the excited-state absorption (Figs. 3a,b). While transient absorption changes are observed for both CuS and Au/CuS films, no transient signals in Au nanoparticle film absorption are detected, since the pump wavelength is not resonant with the Au plasmons, nor do Au NPs strongly absorb at 2400 nm. The risetime we observe here (~ 130 fs) is significantly longer than the risetime of the pump pulse (83.2 fs), which indicates a non-instantaneous transient system response.

The transient kinetics are fit with a double exponential function and show strong evidence for plasmonic energy transfer (Table 1). The risetime corresponds to hole-hole scattering while the recovery time corresponds to hole-phonon scattering (34,35). The risetime is ~150 fs for all the systems containing CuS. However, the relaxation dynamics reveal a 25% faster decay in the Au/CuS heterostructure (690 fs) than in CuS films (929 fs). The faster Au/CuS relaxation can be attributed to the fact that the HIPRET process between CuS and Au provides an energy-decay pathway in addition to hole-phonon scattering present in pure CuS. Dynamics in Au/CuS films with different separation distances are also measured (Fig. 3b). As the inter-layer spacing grows past the effective coupling distance (10 nm), the relaxation time reverts to that of isolated CuS NPs. This is consistent with our observations in harmonic generation measurements, where there is little SHG and THG enhancement beyond that separation distance (Figs. 1c,d). Faster plasmon relaxation from our ultrafast studies strongly indicates the presence of HIPRET between CuS and Au plasmons in the heterostructure films.

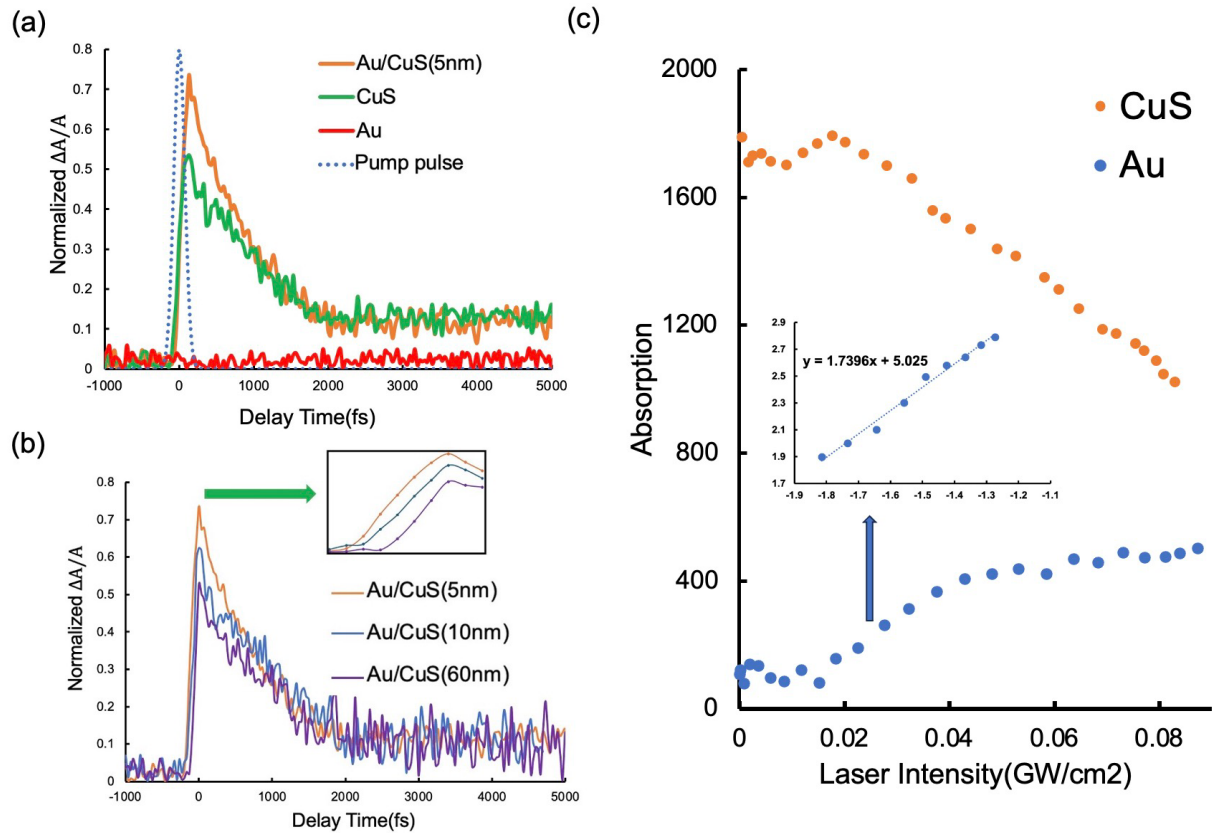

**Figure 3**. Normalized transient absorption dynamics of (a) Au/CuS, CuS and Au. The pump pulse is shown with a dotted-dashed line. (b) Heterostructure Au/CuS films with different separation distances probed at 2400 nm upon 1200 nm photoexcitation at pump fluence of 1.12 mJ/cm$^2$. The inset shows the comparison of the risetime. (c) The absorption of CuS and Au films as a function of the input laser intensity. The inset is the double-logarithmic plots of Au absorption as a function of pump laser intensity above the absorption threshold.

Table 1. Transient Absorption growth and recovery kinetics*.

| System | $\tau_{rise}$ (fs) | $\tau_{recovery}$ (fs) |
|---|---|---|
| Au/CuS(5nm) | 130±12 | 690±3 |
| Au/CuS(10nm) | 132±18 | 937±10 |
| Au/CuS(60nm) | 161±25 | 955±25 |
| CuS | 173±17 | 929±3 |

*Pump: 1200 nm, probe: 2400 nm

**Harmonic-Induced Plasmonic Resonant Energy Transfer**

Thus, we can conclude that, in our dual-plasmonic system, ultrafast bidirectional HIPRET between the Au and CuS NPs serves as the mode of action for enhanced SHG and cascaded THG. High intensity absorption measurements were performed on CuS and Au nanoparticle films individually. (Figure 3c) Absorption bleaching in CuS under high-intensity 1050 nm excitation indicates saturation of the linear absorption, maximizing resonant response of the CuS nanoparticles, whose strong local electric fields contribute to the generation of second-order polarization and thus second harmonic emission. The Au films exhibit a significant increase in absorption as intensity increases. Although Au nanoparticles do not possess high absorption at 1050 nm, this increase in absorption with pump intensity suggests that two-photon absorption, exciting the fundamental mode of their LSPRs, is occurring. However, due to the band structure of CuS—which is indirect with a gap of 2.5 eV—rather than decaying the excited state by exciton formation followed by recombination and two-photon photoluminescence, another competitive process dominates. In this alternative process, with both the second-harmonic of the plasmon mode in CuS and fundamental plasmon resonance in Au sharing spectral overlap, these two plasmonic dipoles can couple to each other, exhibiting the characteristic inverse-sixth power distance sensitivity. The coupled harmonic energy can be radiated as second-harmonic light or participate in sum-frequency generation with another pump-laser photon to enhance third-harmonic generation.

**Plasmonic Couplings for Resonant and Off-Resonant Pumping**

Thus far, it has been assumed that the harmonic relation of the LSPRs is essential to enhanced harmonic generation via the HIPRET process. While the measured distance dependence is consistent with that model, the hypothesis must be tested by detuning the pump wavelength from the peak of the CuS plasmon resonance in both blue and red directions while collecting the upconverted spectra for comparison to the resonant condition.

For these investigations, a new experiment was prepared with an Orpheus-F optical parametric amplifier (OPA) providing excitation wavelengths from 600 nm to 2000 nm (Fig. S11). Experimentally, the pump wavelength is blue-shifted to 800 nm and red-shifted to 1200 nm and 1600 nm and compared to the harmonic yields in the original experimental condition of 1050 nm (Fig. 4). We observe the dominant THG signal at 1050 nm and 1200 nm pump wavelengths, modest THG for 1600 nm pump wavelength, and no measurable THG for 800 nm pump wavelength. A SHG signal less intense than the corresponding THG signal is observed for all pump wavelengths. Moreover, a peak centered at 717 nm is seen under all excitation conditions. The upconverted 717 nm emission is likely from the Au MPPL due to the radiative recombination of the *sp*-band electrons with the *d*-band holes triggered by the sequential steps of the photon absorption, which is consistent with a previous report (36).

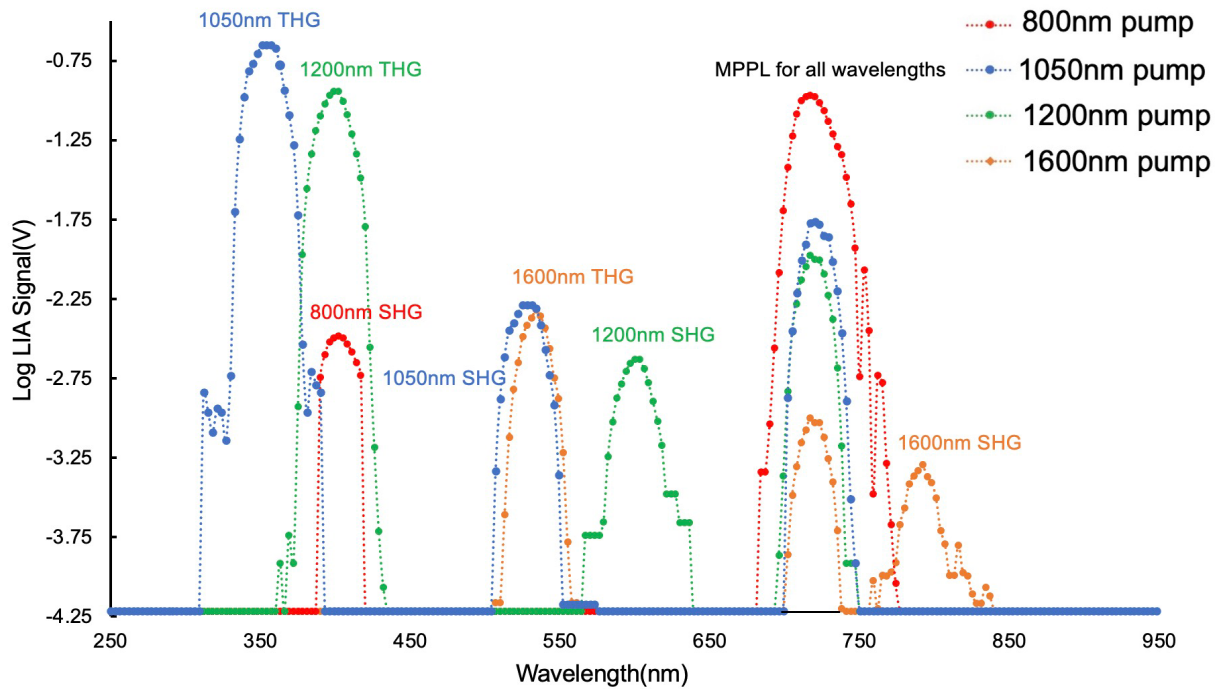

Figure 4. Semi-logarithmic (base 10) plot of the output spectra from Au/CuS heterostructure films (2-3nm ligand separated) under the tunable excitation wavelength: 800 nm (red), 1050 nm (blue), 1200 nm (green), 1600 nm (orange) at the fixed pump intensity 0.08 GW/cm$^2$.

The THG and SHG intensities are plotted as a function of the input laser intensity for different pump wavelengths (Figs. 5a and 5b). The highest third harmonic intensity is seen under 1050 nm pump excitation and is attributed to the enhancement from the cascaded process as the LSPRs of CuS and Au NPs satisfy the fundamental and second harmonic condition of 1050 nm excitation, respectively. When the pump frequency is red-shifted from the resonant condition (1050nm) to 1200 nm, the THG and SHG intensities both decrease. The THG reduces much more than the SHG probably due to the broad LSPR spectrum of CuS NPs in the near-infrared range, which could be excited by either 1050 nm or 1200 nm pump. However, the second harmonic of 1200 nm excitation well lies outside the resonant spectrum of the Au LSPR (525 nm peak). Thus, two photon excitation of the Au LSPR cannot occur. Additionally, without spectral overlap between the second harmonic of the excitation and the Au LSPR, HIPRET cannot occur—thus, harmonically resonant SFG (the second step in cascaded THG) is no longer operative. The efficiency of harmonic generation is then lower than with 1050 nm excitation (which satisfies the resonant condition). In contrast to excitation at 1050 nm and 1200 nm, the SHG and THG from 1600 nm and 800 nm pump wavelength are much smaller, since the fundamental and second harmonic frequency of the pump deviate even farther from the peak of the CuS and Au LSPRs, respectively. As a result, the cascaded THG process is not triggered at all. The nonlinear orders of the measured harmonic generation signals are confirmed by the log-log plots (Figs. 5d and 5e).

The observed photoluminescence signals provide more insight into the upconversion mechanisms with the pump wavelength detuned from the plasmon resonance (Fig. 5c). An extremely large photoluminescence signal under 800 nm pump excitation stands out. The log-log fitting coefficient is close to 1, which suggests that single-photon photoluminescence in Au dominates the upconversion process in the Au/CuS heterostructure under the 800 nm excitation where the SHG and THG enhancement vanish. The 800 nm pump energy matches the interband transition in Au (37) but is not sufficient to trigger excitonic absorption in CuS (2.5 eV, 497 nm) (38). The results also show decreasing MPPL intensity as the excitation red-shifts from 800 nm to 1600 nm. As the pump wavelength increases, the single-photon excitation energy decreases and the number of the photons required for the MPPL process increases, which could be reflected by the increasing slope of the log-log plots (Fig. 5f). Since more photons are required, the electronic transition probability (and MPPL intensity) decreases.

By comparing THG, SHG and MPPL intensities at various pump wavelengths (Figs. 5 and S13), it can be concluded that the harmonic condition between the plasmon resonances of the NPs and the pump wavelength is essential for the efficient plasmonic interaction and enhanced harmonic generation. As the excitation wavelength is red-shifted from the harmonic condition, the intensities of all the upconverted signals (SHG, THG, MPPL) are reduced, while blue-shifting the excitation wavelength switches the dominant upconversion process from harmonic generation to MPPL. Our results thus present the first observation of the effect of resonant and off-resonance plasmon-plasmon coupling on the upconversion mechanism in these multi-plasmonic nanostructures.

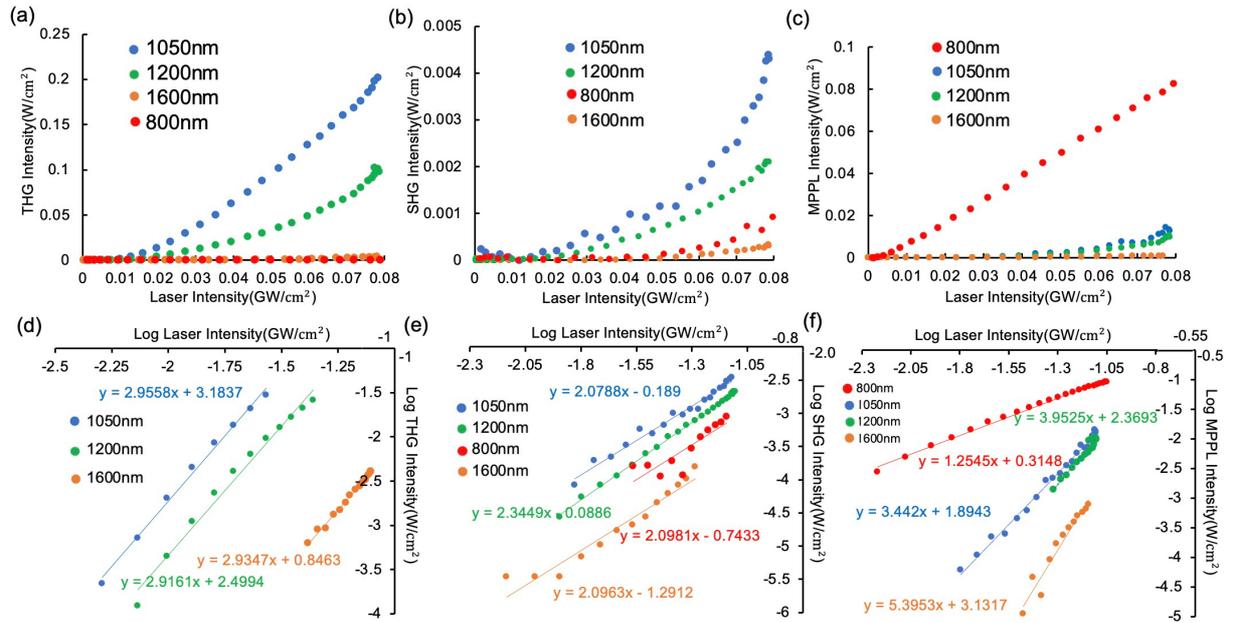

Figure 5. (a) Intensity of the upconverted signal generated by tunable excitation wavelength as a function of the input laser intensity: (a) THG (b) SHG (c) MPPL. Double-logarithmic plots of output intensity as a function of pump laser intensity: (d) THG (e) SHG (f) MPPL.

## Discussion

Previously we reported that the yields of second-harmonic and cascaded third-harmonic emission were enhanced in dual-plasmonic Au/CuS films by factors of 3.3 and 20 compared to the incoherent sum from the constituent NPs alone (30,31), but control over the separation distance between the plasmonic films was limited and the mechanism by which the plasmonic interactions incorporate into these enhanced nonlinear processes remained ambiguous. Here, we observed that dual-plasmonic Au/CuS heterostructures produce greatly enhanced second-harmonic and cascaded third-harmonic generation *via* resonant plasmon-plasmon coupling between Au LSPRs and the second harmonic of CuS surface plasmons. By depositing alumina layers of various thicknesses, the separation distance between the Au and CuS layers is precisely manipulated. This separation distance correlated with the inverse-sixth power dependence of harmonic intensities on pump intensity, and experimental measurements are corroborated by FDTD simulation in Lumerical®. Transient absorption spectroscopy shows much faster plasmon relaxation dynamics in heterostructures than in pure CuS nanocrystals, due to transfer of plasmonic energy between CuS and Au NPs. These results demonstrate a novel type of plasmonic interactions– –HIPRET—through which the second-harmonic frequency components generated from the CuS plasmon couples resonantly with the nearby fundamental Au dipolar plasmon modes, resulting in the enhancement of the harmonic emission from both plasmonic counterparts in the heterostructure. Furthermore, the resonant harmonic condition between the nanoparticulate LSPRs and the pump wavelength is investigated. The red-shifted pump wavelength attenuates the upconversion efficiency, while the blue-shifted excitation energy changes the dominant upconversion process from harmonic generation to MPPL.

Our comprehensive study of the fundamental mechanism underlying the plasmon-plasmon interactions in Au/CuS dual-plasmonic nanoparticle heterostructure paves the way for designing a new generation of ultrafast, high-efficiency thin-film nonlinear optical nano-devices. The remarkable tunability of the plasmon resonance in these metal-semiconductor nanosystems — achieved by varying the shape, size, composition and doping level of the semiconductor nanoparticles — enables a broad frequency spectrum for both light absorption and upconversion. Additionally, the wide range of possible combinations of the plasmonic metals and semiconductors allows the system to be explored for diverse applications, including spectroscopy, telecommunications, photocatalysis, quantum information, biosensing and biomedical imaging. One intriguing example is that the high harmonic generation demonstrated in our previous work (29) could provide multi-spectral probes of thin layered materials deposited on top of multi-plasmonic layered nanoparticle structures (39).

Table S1 lists several well-studied plasmonic metal and semiconductor materials along with the wavelengths of their plasmon resonances. These dual-plasmonic heterostructures can be investigated for other ultrafast second-order nonlinear processes. For example, by carefully choosing the plasmon resonances of the metal and semiconductor to match the signal (in visible range) and idler (in the NIR range) wavelength of optical parametric amplifier, light at telecom wavelengths (1550nm) could be efficiently generated by plasmon-enhanced difference frequency generation. Efficient generation of single-photon pairs by spontaneous parametric down-conversion is another possible application, which might lead to quantum information and computation. In metal-semiconductor nanocomposites, by matching the metal and semiconductor LSPRs at the pump and signal frequencies, respectively, single-pair generation rates are expected to be significantly enhanced.

## Materials and Methods

Nanoparticle Synthesis and Film Deposition

Nanoparticles of CuS and Au are synthesized using the standard solvothermal technique and then assembled into heterostructure films sequentially by spin coating as described in previous reports (40,41). The alumina of varying thickness is vertically deposited *via* electron-beam physical vapor deposition (Angstrom Amod) onto the Au layers after the Au film deposition on the glass slides. The electron beam evaporation is carried out at the pressure $5 \cdot 10^{-6}$ Torr and the electron beam is rastered across the alumina piece precursor in a crucible while the power increases until the deposition rate stabilizes at 0.3 Å/s, at which point the sample shutter opens and the deposition on the sample initiates. Finally, the CuS NPs will be added atop the alumina layers *via* sputter coating. The thickness of the deposited alumina is verified by atomic force microscopy (AFM) (Bruker Dimension Icon AFM) in the tapping mode. The structure of the alumina deposition on the Au films is confirmed with scanning electron microscopy (SEM) (Zeiss Merlin SEM) at 1.20 kV and 10.00 kV with the high-efficiency secondary electron detector and energy dispersive spectroscopy at 10.00 kV and 500 pA (Oxford X-MAX 50 SDD and Oxford Aztec 3.3 EDS analysis software).

Nonlinear Optical Measurements

The extinction spectrum of the Au-Al$_2$O$_3$-CuS films is acquired in the Agilent Technologies Cary 5000 UV-vis-NIR spectrophotometer with an integrating sphere from 400 nm to 3000 nm. Optical measurements are performed using the nonlinear setup depicted in Fig. S12. The Orpheus-F OPA is pumped by a compact, high repetition-rate femtosecond Pharos laser comprising an oscillator and chirped-pulse amplifier, employing diode-pumped Yb:KGW as an active medium and generating pulses centered at $1035\pm5$ nm at an average power of 600 mW, 187 fs duration, at a repetition rate of 60 kHz and an energy per pulse of 10 µJ. The signal and idler pulses of the OPA are tunable over 650-1000 nm and 1000-2500 nm ranges, respectively, providing gap-free tunability. The laser beam from the OPA is mechanically chopped at a frequency of 265 Hz, with a duty cycle of 20%. This avoids repetitive exposures and ensures stable measurement conditions. To determine the dependence of the upconverted signal on the pump intensity, we use a rotational polarizer to adjust the laser power by varying the angle between the fixed output pump polarization of the pump laser and the axis of the rotational polarizer. Pump power is measured in a Thorlabs S130C power meter with a PM100D readout. The SHG, THG and MPPL signals are isolated from other upconverted signals using the respective band pass filter for each pump wavelength (Optosigma VPF-25C-10-25-350 for 1050 nm THG, Optosigma YIF-BA515-560S for 1050 nm SHG and 1600 nm THG, Optosigma VPF-25C-10-12-265 for 800 nm THG, Thorlabs FBH400-10 for 800 nm

SHG and 1200 nm THG, Thorlabs FBH600-10 for 1200 nm SHG and Thorlabs FBH800-10 for 1600 nm SHG). Our detector is the solid-state photomultiplier tube (PMT, Hamamatsu, R9875U for ultraviolet light detection and R9880U for visible light detection) operating at 1.1 kV. A Newport ¼ m 74100 Monochromator and the PMT detector are used to collect the output spectrum of the nanoparticle heterostructures.

Ultrafast pump-probe measurement

The ultrafast dynamics of nanoparticle films are measured using the signal (105 fs, 1200 nm) and idler (130 fs, 2400 nm) beam from a Light Conversion Topas OPA pumped by a 1kHz Ti:sapphire regenerative amplifier (Spectra Physics Spitfire Ace, 100 fs, 800 nm) (42). The signal (pump) is chopped at 500Hz while the idler (probe) retains a 1kHz repetition rate. The pump is delayed relative to the probe by computer-controlled delay stage. To ensure probing of a uniformly pumped region, the signal and idler beams are focused to beam waists of 200 $\mu$m and 90 $\mu$m, respectively, measured *via* knife edge experiments. A linear polarizer on a motorized rotational mount adjusts the pump fluence at 1.12 mJ/cm$^2$. The probe is attenuated using a neutral density filter (ND3) and a linear polarizer to achieve an incident fluence of less than 24 $\mu$J/cm$^2$. The transmitted probe signal is detected using a PbS fixed gain detector (Thorlabs PDA30G) *via* an SR830 lock-in amplifier (LIA) referenced by the chopper. The transient kinetic traces are plotted with 5 points moving average smoothing. To determine the rate of thermalization due to hole-hole scattering and relaxation due to hole-phonon scattering, the rise-time of the transient absorption change is fit to a phenomenological response function (43):

$$u(t) = H(t)\left[1 - exp\left(-\frac{t}{\tau_{\text{rise}}}\right)\right] exp\left(-\frac{t}{\tau_{\text{recovery}}}\right) \qquad (4)$$

and the decay trace is fit to the mono-exponential:

$$u(t) = A * exp\left(-\frac{t}{\tau_{\text{recovery}}}\right) \qquad (5)$$

where *H(t)* is the Heaviside step function, $\tau_{\text{rise}}$ and $\tau_{\text{recovery}}$ are the time constants of hole-hole scattering and hole-phonon scattering process.

Optical simulations

The theoretical simulations of the absorption spectra and near-field distributions are performed by commercially available software (Ansys Lumerical® FDTD 2021 R2) using a finite-difference time-domain (FDTD) solver. The material properties of Au sphere (15 nm diameter) are taken from the Johnson and Christy dataset (44), while the CuS polygon (base side length 7.3 nm, 6.7 nm height) are created as Drude plasma model. The damping constants and plasma frequency of these models are adopted from the reference (41). The plane wave centered at 1050 nm is used as the incident pump. The perfectly matched layer (PML) boundary conditions are employed in all x, y and z directions to prevent any non-physical scattering at boundaries. The grid size in all axes is set to 0.6 nm for the overall simulated region. The simulation time is set to 1000 fs and the auto shutoff min is set to $10^{-5}$ s. The absorption cross-section is computed by a power monitor and the near-field patterns in the coupling regions between the Au and CuS NP are examined by a frequency domain field profile monitor.

## Acknowledgements

Y.Y and A. J. C. are thankful to the Vanderbilt Department of Physics and Astronomy for graduate assistantships.

## Author information

### Contributions

Y. Y, and N. J. S. conceived this project. Y. Y., N. J. S and A. J. C. synthesized the nanoparticles, deposited and characterized the films. A. J. C. and Y. Y. performed the pump-probe experiments. Y. Y. conducted the nonlinear optical measurements, carried out the FDTD simulation and developed the theoretical expressions ー with critical input and supervision by J. E. M. and R. F. H. Y. Y. wrote the original paper and revised the manuscript with help of N. J. S., A. J. C., K. K., J. E. M. and R. F. H. All authors have given approval to the final version of the manuscript.

### Corresponding author

Correspondence to Richard F. Haglund

## Ethics declarations

Conflicts of interest

The authors declare no conflicts of interest.

## Supplementary Information

**This PDF file include:**

Supplementary Figure S1-S14, Table S1 and theoretical derivation of harmonic-induced plasmonic resonant energy transfer in Section S2.

Supplementary Materials for:

# Harmonic-Induced Plasmonic Resonant Energy Transfer between Metal and Semiconductor Nanoparticles

Yueming Yan *et al*

Corresponding Author: Richard F. Haglund*

E-mail: richard.haglund@vanderbilt.edu

**This PDF file include:**

Figures S1-S14

Table S1

Supplementary Text: Section S2. Harmonic-Induced Plasmonic Resonant Energy Transfer (HIPRET) in Au/CuS Heterostructure Films for SHG, THG, and FDTD simulation

**Supplementary Figures**

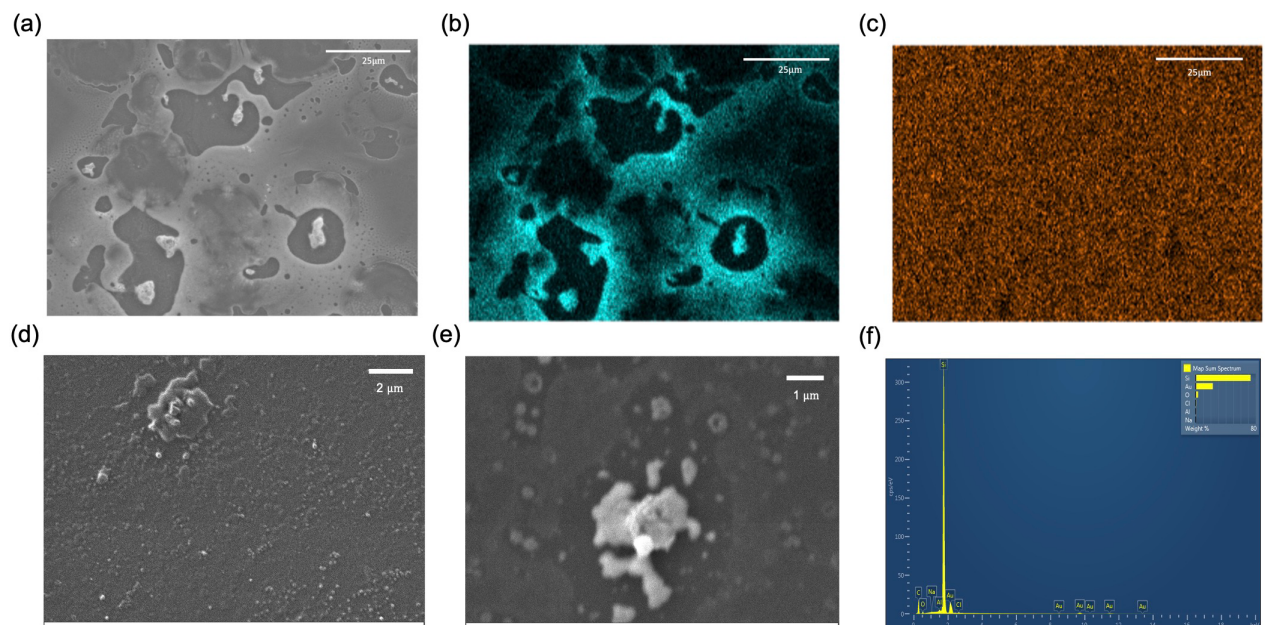

Figure S1. (a) SEM image of $Al_2O_3$ deposited on the Au films. (b) Elemental mapping of Au in (a). (c) Elemental mapping of Al in (a). (d) and (e) SEM images with increased magnification collected with type-II secondary electrons (SE2) that are generated when a backscattered electron leaves the surface. (f) The energy dispersive spectroscopy (EDS) spectrum of the region in (a).

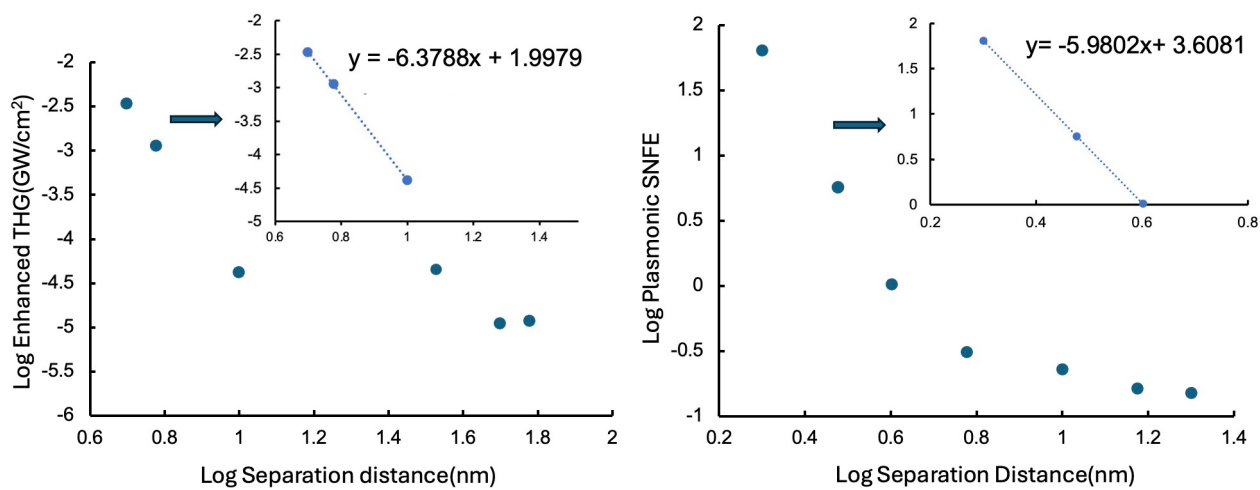

Figure S2. Double-logarithmic plots of (a) enhanced THG and (b) plasmonic square of near-field enhancement (SNFE) as a function of separation distance dependence. The enhanced THG and plasmonic SNFR are obtained by substracting the incoherent sum between CuS and Au NPs. The insets represent the linear fit of the first three points. The results strongly support the inverse-sixth power distance -dependence of plasmonic interaction: within the demonstrated effective distance, the fitting coefficienct of log-log plot is nearly -6 for both the experiment and simulation.

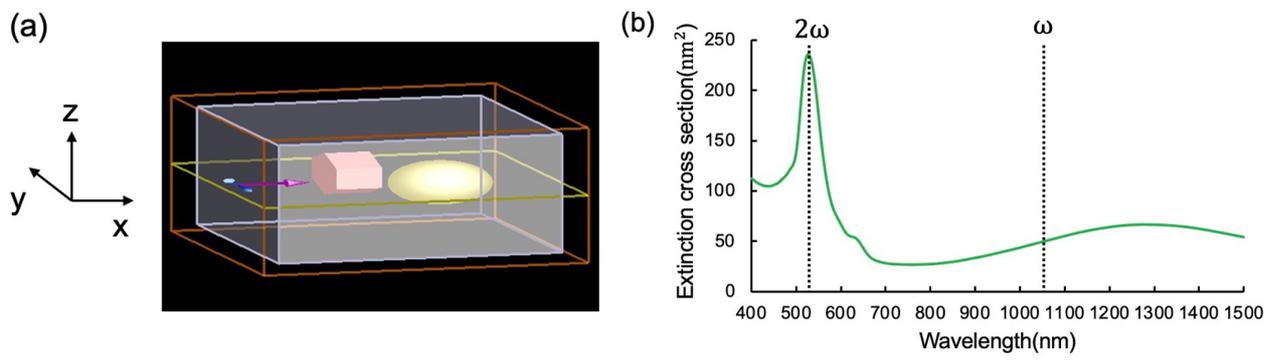

Figure S3. (a) Structural model of Au/CuS nanoparticles in the FDTD simulation from the perspective view. The red arrow indicates the propagation of a plane wave input light, and the blue arrow indicates the polarization. (b) The simulated extinction cross-section.

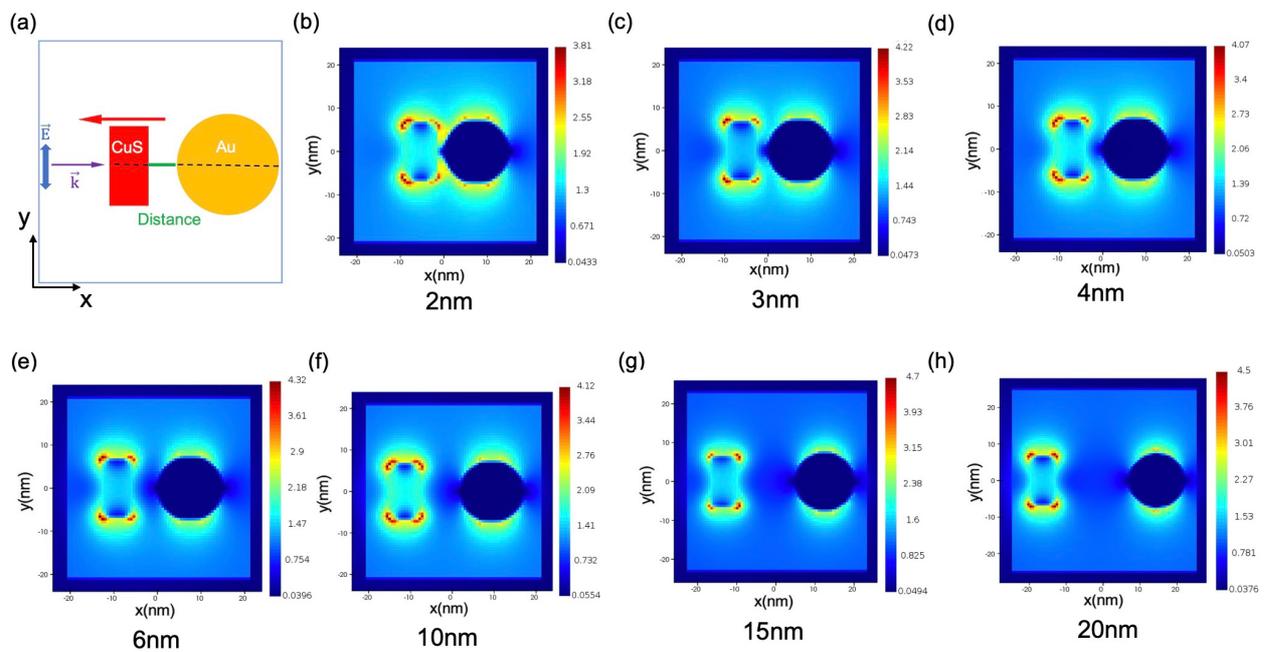

Figure S4. (a) Structural model of simulated Au/CuS nanoparticles when there is no offset between Au and CuS along the y axis. The Au and CuS are moved along the x axis (red arrow) to change the separation distance. FDTD calculated local electric field distribution with varying separation distance between the Au and CuS nanoparticle (b) 2 nm (c) 3 nm (d) 4 nm (e) 6 nm (f) 10 nm (g) 15 nm (h) 20 nm. The color scale bars show the relative increase in the electric field $|E/E_0|$.

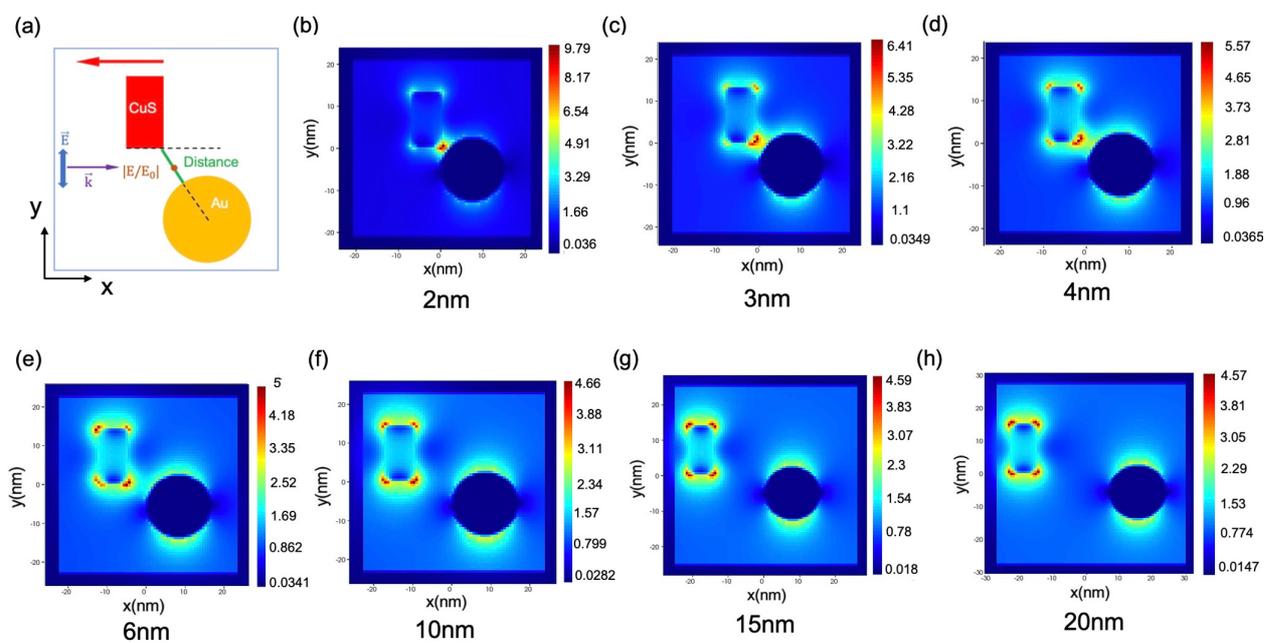

Figure S5. (a) Structural model of simulated Au/CuS nanoparticles when the nanoparticles are moved along the x axis to change the separation distance (red arrow) with a fixed y-axis offset. FDTD calculated local electric field distribution with the varying separation distance between the Au and CuS nanoparticle (b) 2 nm (c) 3 nm (d) 4 nm (e) 6 nm (f) 10 nm (g) 15 nm (h) 20 nm. The color scale bars show the relative increase in the electric field $|E/E_0|$.

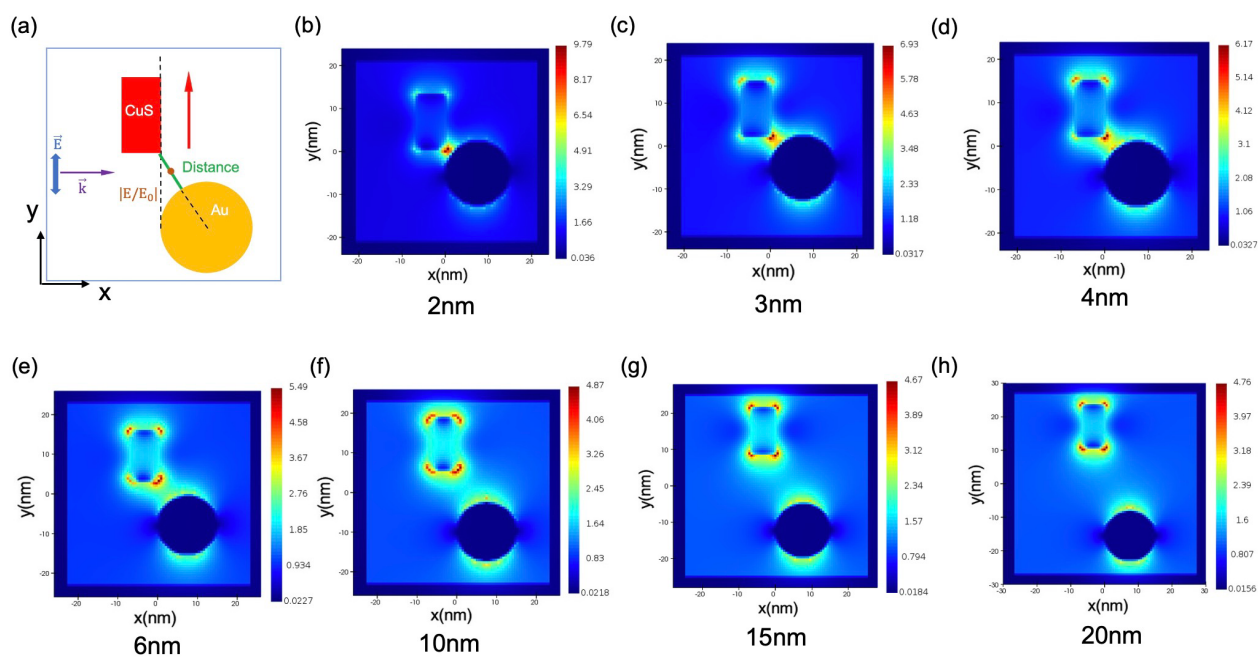

Figure S6. (a) Structural model of simulated Au/CuS nanoparticles when the nanoparticles are moved along the y axis to change the separation distance (red arrow) with a fixed x-axis offset. FDTD calculated local electric field distribution with varying separation distance between the Au and CuS nanoparticle (b) 2 nm (c) 3 nm (d) 4 nm (e) 6 nm (f) 10 nm (g) 15 nm (h) 20 nm. The color scale bars show the relative increase in the electric field $|E/E_0|$.

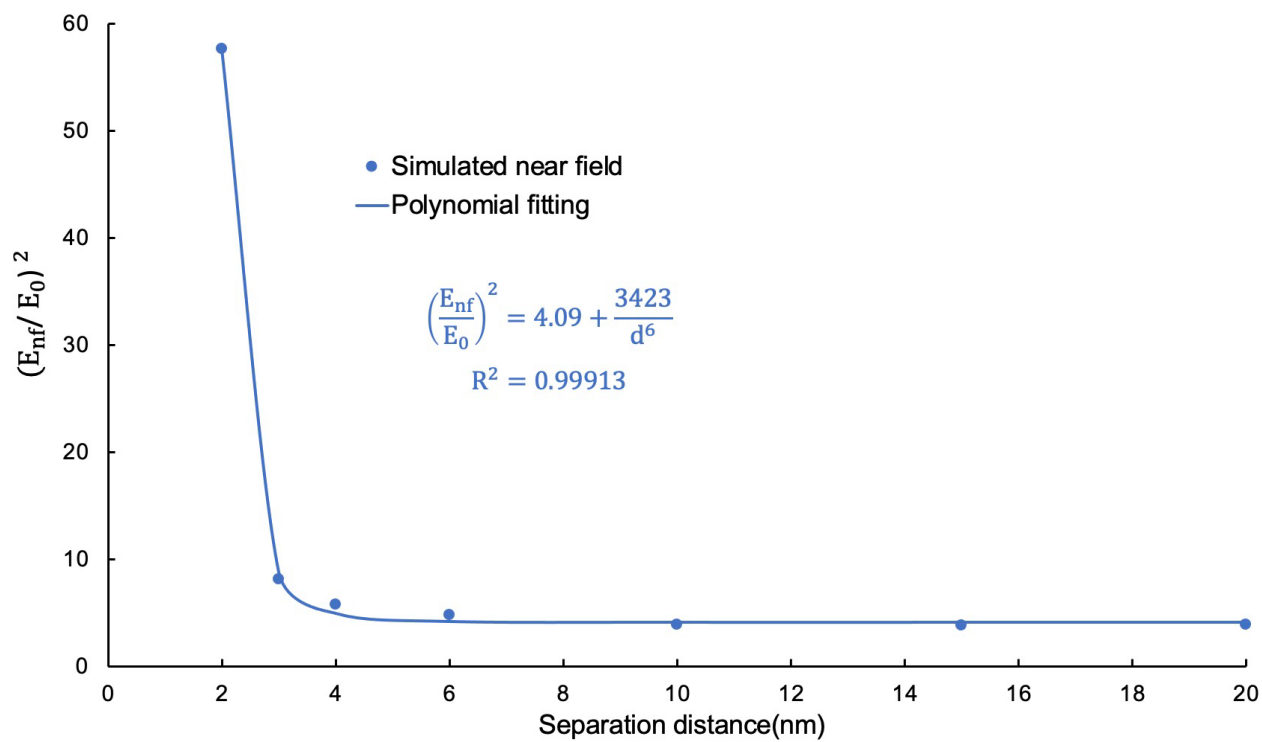

Figure S7. The statistic average of the squared near field enhancement between Figure S5 and S6 as a function of the separation distance (points) with the inverse-sixth polynomial fit (smooth curve). For each separation distance, the squared field enhancement is recorded at the same position: 1 nm from the Au surface on the line connecting the edge of CuS and Au center. (Brown point in Figure S5a and S6a).

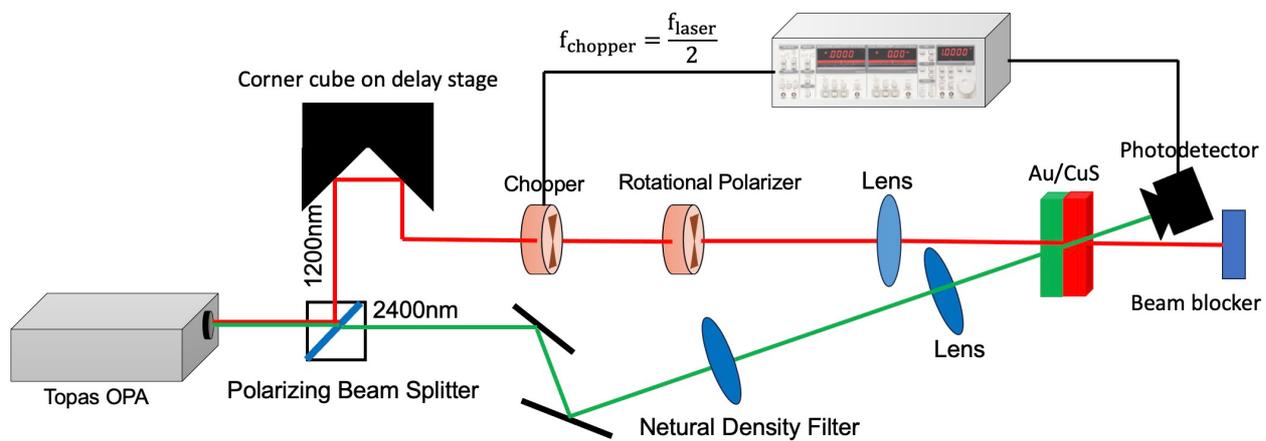

Figure S8. Simplified diagram of the optical setup for ultrafast pump-probe measurements.

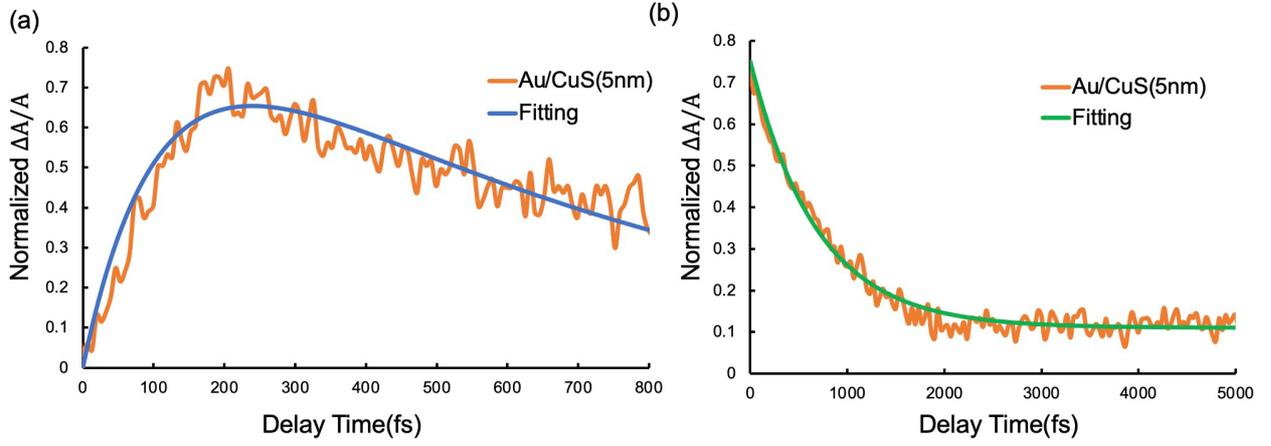

Figure S9. The experimentally measured (a) risetime and (b) decay time together with fits of normalized transient absorption dynamics of Au/CuS(5nm) by

$$u(t) = H(t)\left[1 - exp\left(-\frac{t}{\tau_{rise}}\right)\right] exp\left(-\frac{t}{\tau_{recovery}}\right) \quad (Sa)$$

and

$$u(t) = A exp\left(-\frac{t}{\tau_{recovery}}\right) \quad (Sb),$$

separately. The fitted time constants are: $\tau_{rise}$=130±12 fs, $\tau_{recovery}$=690±3 fs.

Covellite (CuS) nanoparticles studied here are p-type plasmonic materials with lower carrier dentisty ($N_h \approx 10^{22}$ cm$^{-3}$) than noble metals ($N_h \approx 10^{23}$ cm$^{-3}$). The covellite lattice exhibits the inherent metallic-like character arising from the significant density of valence-band free holes associated with the $3p$ orbitals of sulfur. In this way, the holes in CuS can be treated as fully free and the LSPR response and dynamics of CuS can be reasonably well described by the Drude-Sommerfeld and two-temperature models.

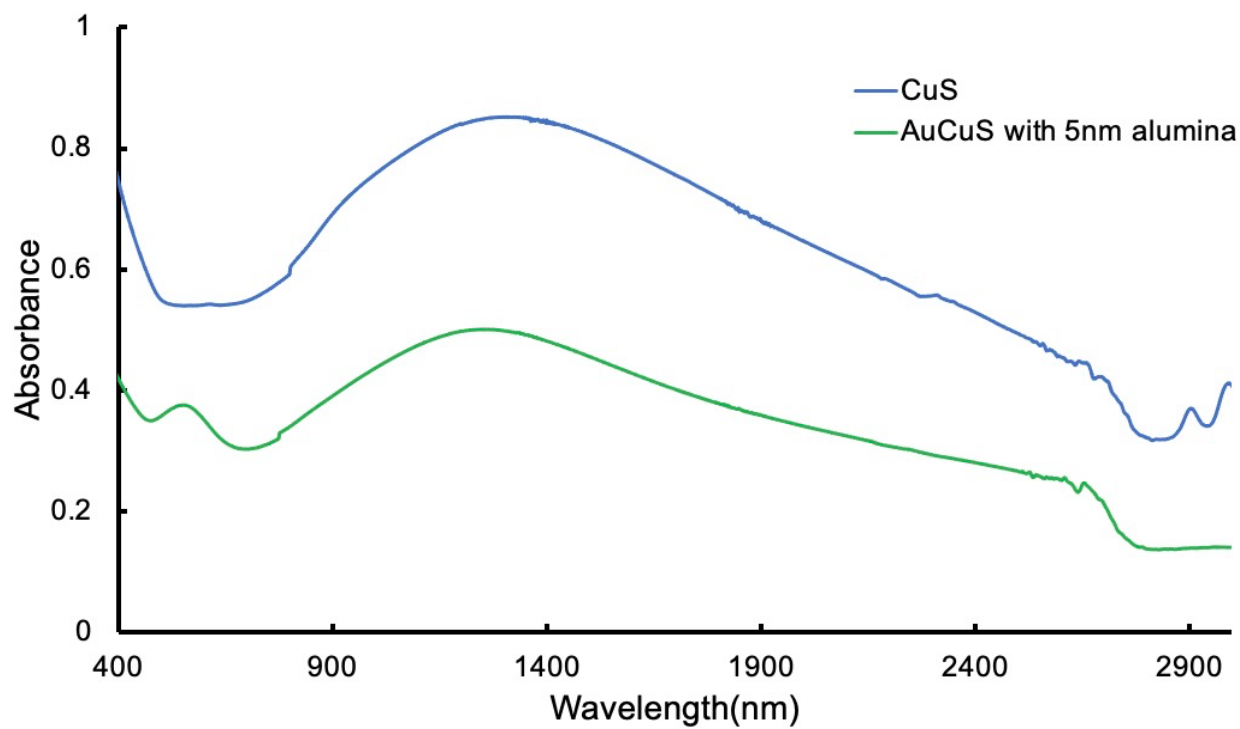

Figure S10. Spectrophotometric absorbance of the pure CuS and Au/CuS with 5 nm alumina layer from 400 nm to 3000 nm.

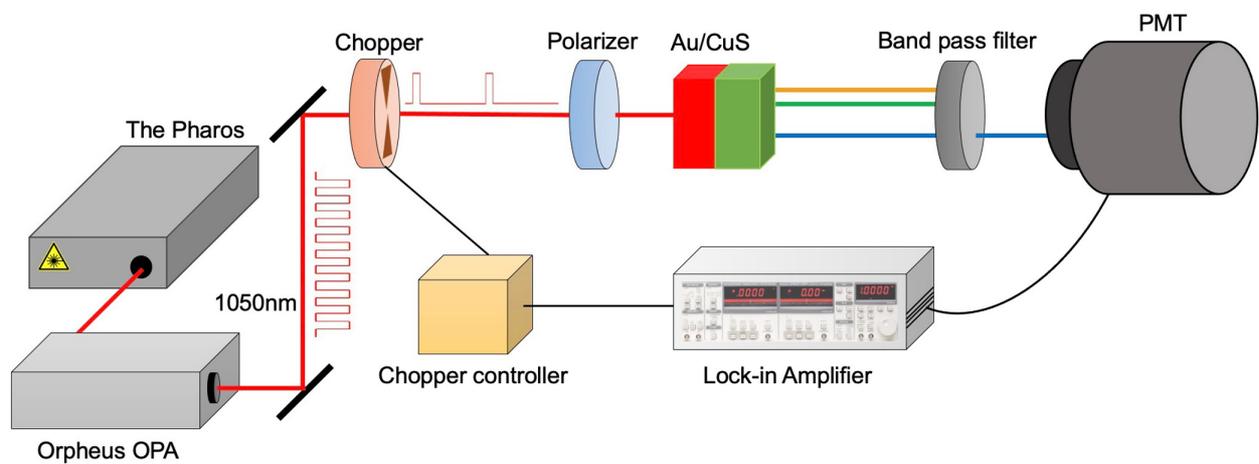

Figure S11. Simplified diagram of the experimental single-beam setup for measuring nonlinear optical response using a tunable OPA system. Different colors correspond to different wavelengths. Red: 1050 nm fundamental beam. Orange: 717 nm MPPL output. Green: 525 nm 2ω beam. Blue: 350 nm 3ω beam.

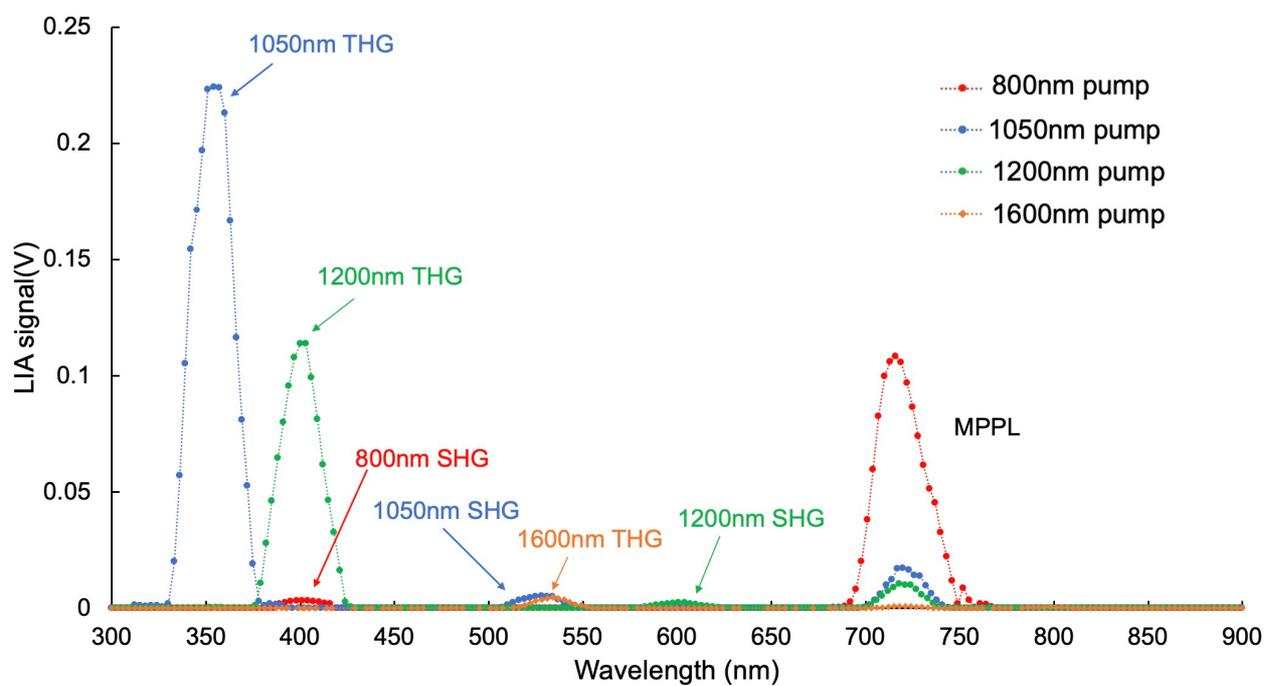

Figure S12. Lock-in amplifier signal from light produced by the Au/CuS heterostructure films under tunable excitation at 800 nm (red), 1050 nm (blue), 1200 nm (green). and 1600 nm (orange).

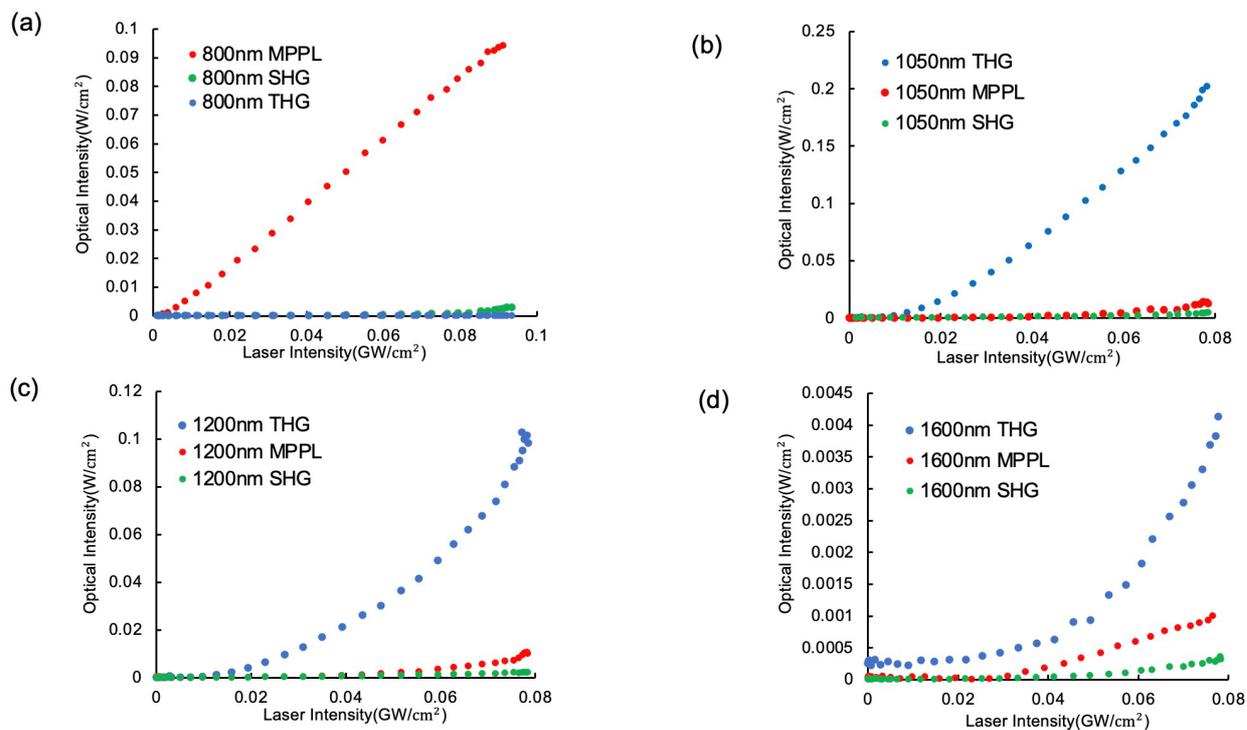

Figure S13. Comparison of the upconverted signal intensity as a function of the pump laser intensity under tunable excitation wavelengths (a) 800 nm pump (b) 1050 nm pump (c) 1200 nm pump (d) 1600 nm pump.

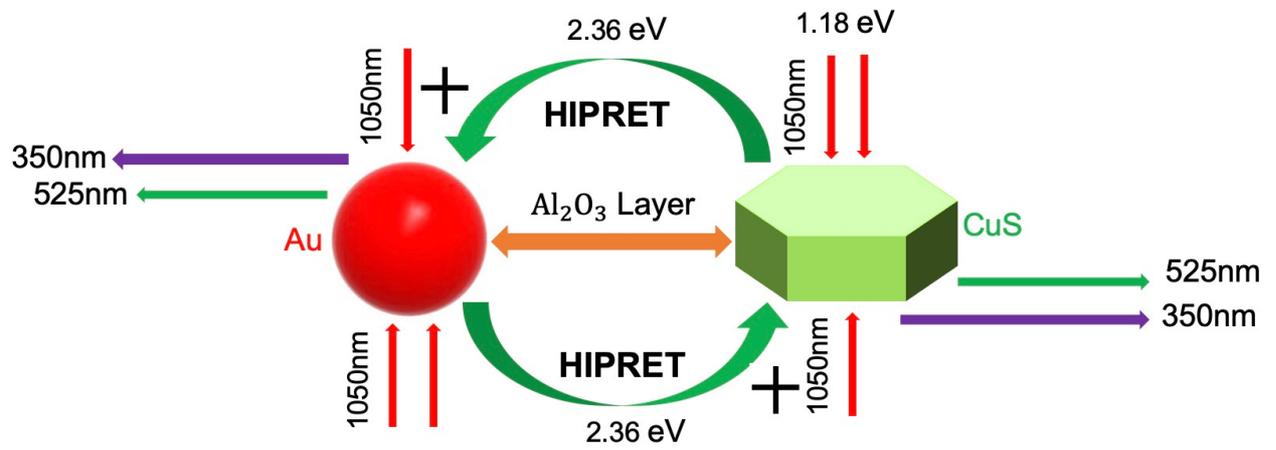

Figure S14. A schematic diagram of the enhanced third-harmonic generation and second-harmonic generation mediated by the bidirectional harmonic-induced resonant energy transfer in the Au/CuS nanoparticle heterostructure.

Table S1. The tunable plasmon-resonance wavelength of some well-studied plasmonic metal and semiconductor nanocrystals.

| Materials | Plasmon resonance (nm) |
|---|---|
| Aluminum | 100-200 |
| Chromium | 200-300 |
| Silver | 320-400 |
| Lithium | 400-450 |
| Gold | 500-600 |
| ZnO | 370-400 |
| Titantium Nitride | 400-1100 |
| Copper Sulfide | 800-3000 |
| Indium Tin Oxide | 1500-2000 |
| Aluminum-doped Zinc Oxide | 2500-4000 |

**Section S2. Harmonic-Induced Plasmonic Resonant Energy Transfer (HIPRET) in Au/CuS Heterostructure Films for SHG, THG, and FDTD simulation**

We measured the intensity of harmonic generation and calculated the square of near-field enhancement to describe the plasmon-plasmon coupling efficiency between CuS and Au plasmons. Here, we develop theoretical expressions for the quantities measured in experimental and theoretical conditions, demonstrating the same inverse-sixth power distance dependence as the resonant energy transfer rate between CuS and Au plasmons. This model provides a deeper understanding of the upconversion mechanism mediated by plasmon-plasmon coupling in multi-plasmonic nanostructures.

**2.1 Second Harmonic Generation (SHG) in Au/CuS Films**

As observed in the intensity-dependent absorption of CuS and Au films, the 1050nm pump laser used in our experiments is strong enough to excite both intensity-dependent second-harmonic components of the plasmon resonance in CuS and the Au LSPR. In harmonic-induced resonant energy transfer, spectral overlap exists between the second harmonic of the CuS LSPR and the fundamental Au LSPR; these dipolar resonances can then couple to each other. The resonant energy transfer afforded by this coupling is bidirectional, leading to enhanced second-harmonic generation (SHG) in both components of the heterostructure films.

First, we consider the contribution of the resonant energy transfer from CuS to Au in enhancing SHG. In this pathway, second-harmonic dipoles in CuS are induced by high intensity exposure to the pump laser. These dipoles then couple to the Au plasmon resonance and energy is transferred. As a result, there are two terms of second-order polarization contributing to the generation of second harmonic light in the Au films:

$$P_{Au}^{(2)} = P_{pump}^{(2)} + P_{CuS-RET}^{(2)} = \varepsilon_0 \chi_{Au}^{(2)} E_{pump}(\omega)^2 + \varepsilon_0 \chi_{Au}^{(1)} E_{CuS-RET}(2\omega) \qquad (S1)$$

The first term, $P_{pump}^{(2)}$ is polarization induced by the fundamental pump field $E_{pump}(\omega)$ via two photon absorption while the second term, $P_{CuS-RET}^{(2)}$ is induced by the CuS near field—$E_{CuS-RET}(2\omega)$. $\chi_{Au}^{(1)}$ and $\chi_{Au}^{(2)}$ are the linear susceptibility and second-order nonlinear optical susceptibility, respectively. The total SHG intensity emitted from Au nanoparticles is thus:

$$I_{Au}(2\omega) = I_{pump}^{Au}(2\omega) + I_{CuS-RET}(2\omega) \qquad (S2)$$

With

$$I^{Au}_{pump}(2\omega) = \frac{8{d^{Au}_{eff}}^2 \omega^2 I(\omega)^2 L_{Au}^2}{n_2^{Au}(n_1^{Au})^2 \epsilon_0 c^3} \qquad (S3)$$

and

$$I_{CuS-RET}(2\omega) = \frac{1}{2} c\epsilon_0 E_{CuS-RET}(2\omega)^2 \qquad (S4)$$

Where $I(\omega)$ is the pump laser intensity, $d^{Au}_{eff}$ is the effective second-order susceptibility tensor for Au films in SHG, $\epsilon_0$ is the permittivity of free space, $c$ is the speed of the light in a vacuum, $L_{Au}$ is the Au film thickness, and $n_2^{Au}$, $n_1^{Au}$ are the refractive index of the Au films at the second harmonic and fundamental frequency, respectively. $I^{Au}_{pump}(2\omega)$ refers to the SHG intensity from the Au nanoparticles induced by the pump laser, while $I_{CuS-RET}(2\omega)$ refers to the intensity of second harmonic light emission induced by energy transfer from CuS. For the ultra-thin films in our work, the phase matching factor is not contained in equation (S3). The plasmon-coupled resonant energy transfer rate from CuS to Au is given by (33):

$$W_{CuS-RET}(\omega) = \frac{\eta_{CuS}}{\tau_{CuS}} \cdot \frac{9c^4}{8\pi} \cdot \left|\frac{\overrightarrow{E_{CuS-RET}}}{\mu_{CuS}}\right|^2 \cdot \int d\omega \frac{\epsilon_{Au}(\omega) F_{CuS}(\omega)}{(\omega)^4} \qquad (S5)$$

where $\eta_{CuS}$ is the conversion efficiency of second-harmonic generation in CuS nanoparticles, $\tau_{CuS}$ is the CuS plasmon lifetime, $\mu_{CuS}$ is the transition dipole of the CuS plasmons, and $F_{CuS}(\omega)$ and $\epsilon_{Au}(\omega)$ represent the plasmon resonance spectra of CuS and Au, respectively. $\left|\frac{\overrightarrow{E_{CuS-RET}}}{\mu_{CuS}}\right|^2$ is the coupling factor, which accounts for the dependence of the coupling on the relative orientations of the Au transition dipole and the electric field induced by CuS's transition dipole, as well as the separation distance. The coupling factor can be written as:

$$\left|\frac{\overrightarrow{E_{CuS-RET}}}{\mu_{CuS}}\right|^2 = \frac{\kappa^2_{CuS-Au}}{\epsilon_r^2 r^6} \qquad (S6)$$

where the $\kappa^2_{CuS-Au}$ is the orientation factor, and r is the separation distance between the CuS and Au nanoparticles. By substituting equation (S6) into equation (S4), therefore, $I_{CuS-RET}(2\omega)$ can be rewritten to:

$$I_{CuS-RET}(2\omega) = \frac{c\epsilon_0 \mu_{CuS}^2 \kappa^2_{CuS-Au}}{2\epsilon_r^2 r^6} = \frac{A_{CuS-Au}}{r^6} \qquad (S7)$$

where $A_{CuS-Au}$ is a coefficient of coupling when second-harmonic energy is transferred from CuS to Au.

Similarly, second-harmonic energy can also be transferred from Au plasmonic dipoles excited by two photon absorption of the pump laser to the CuS plasmons. In this case, the SHG intensity emitted from CuS nanoparticles can be expressed as:

$$I_{CuS}(2\omega) = I_{pump}^{CuS}(2\omega) + I_{Au-RET}(2\omega) \tag{S8}$$

Specifically, the first term, $I_{pump}^{CuS}(2\omega)$ describes the SHG intensity from CuS nanoparticles induced by the pump laser:

$$I_{pump}^{CuS}(2\omega) = \frac{8 d_{eff}^{CuS^2} \omega^2 I(\omega)^2 L_{CuS}^2}{n_2^{CuS}(n_1^{CuS})^2 \epsilon_0 c^3} \tag{S9}$$

where $d_{eff}^{CuS}$ is the effective second-order susceptibility tensor for CuS films in SHG, $L_{CuS}$ is the CuS film thickness, and $n_2^{CuS}$, $n_1^{CuS}$ are the refractive index of the CuS films at the second harmonic and fundamental frequency, respectively. The second term, $I_{Au-RET}(2\omega)$, represents the intensity of second harmonic light emitted from CuS induced by energy transfer from the Au and is expressed as:

$$I_{Au-RET}(2\omega) = \frac{1}{2} c\epsilon_0 E_{Au-RET}(2\omega)^2 \tag{S10}$$

Substituting the expression for $E_{Au-RET}(2\omega)^2$ into equation (S10) gives:

$$I_{Au-RET}(2\omega) = \frac{c\epsilon_0 \mu_{Au}^2 \kappa_{Au-CuS}^2}{2\epsilon_r^2 r^6} = \frac{A_{Au-CuS}}{r^6} \tag{S11}$$

The total SHG intensity from the heterostructure Au/CuS films can be expressed as:

$$I_{Au/CuS}(2\omega) = [I_{pump}^{Au}(2\omega) + I_{pump}^{CuS}(2\omega)] + \frac{A_{CuS-Au} + A_{Au-CuS}}{r^6} \tag{S12}$$

The total SHG from the hybrid films consists of two components: the incoherent sum of SHG from Au and CuS nanoparticles without plasmonic coupling (directly generated from stimulation by the laser), and the SHG induced by the resonant energy transfer, which shows a characteristic inverse-sixth power dependence of the dipole-dipole interaction.

## 2.2 Third Harmonic Generation (THG) in Au/CuS Films

The enhanced third harmonic generation (THG) in Au/CuS films due to plasmonic interaction, results from a cascaded process where the coupled second harmonic energy combines with pump-laser photons to generate third harmonics through sum-frequency generation (SFG) (31). In addition to contributing to radiation of second-harmonic light, the second-order polarization of Au and CuS nanoparticles can combine with another photon of fundamental harmonic light from the laser to generate the third-harmonic.

Let us first consider the THG in Au nanoparticles. Based on the expressions for SFG, the cascaded third harmonic intensity takes the following form (45):

$$I_{Au}^{Cascaded}(3\omega) = \frac{18 d_{eff}^{Au'^2} \omega^2 I(\omega) I_{Au}(2\omega) L_{Au}^2}{n_3^{Au} n_2^{Au} n_1^{Au} \epsilon_0 c^3} = B_{Au} I(\omega) I_{Au}(2\omega) \quad (S13)$$

where $d_{eff}^{Au'}$ is the effective second-order susceptibility tensor for Au films in SFG, $n_3^{Au}$ is the refractive index of the Au films at the third harmonic frequency, and for simplicity, all factors except for intensity have been combined into a single coefficient $B_{Au}$. By substituting the equation (S2) and (S7) into (S13), the $I_{Au}(3\omega)$ is given by:

$$I_{Au}^{Cascaded}(3\omega) = B_{Au} I(\omega) I_{pump}^{Au}(2\omega) + \frac{A_{CuS-Au} B_{Au} I(\omega)}{r^6} \quad (S14)$$

The total THG produced by Au films consists of cascaded THG and direct THG:

$$I_{Au}(3\omega) = I_{Au}^{direct}(3\omega) + B_{Au} I(\omega) I_{pump}^{Au}(2\omega) + \frac{A_{CuS-Au} B_{Au} I(\omega)}{r^6} \quad (S15)$$

where

$$I_{Au}^{direct}(3\omega) = \frac{27 d_{eff}^{Au(3)^2} \omega^2 I(\omega)^3 L_{Au}}{n_3^{Au} (n_1^{Au})^3 c^3} \quad (S16)$$

Here $d_{eff}^{Au(3)}$ is the effective nonlinear third-order tensor for Au films in direct THG.

Similarly, the third-harmonic intensity produced by CuS films can be expressed as:

$$I_{CuS}(3\omega) = I_{CuS}^{direct}(3\omega) + B_{CuS} I(\omega) I_{pump}^{CuS}(2\omega) + \frac{A_{Au-CuS} B_{CuS} I(\omega)}{r^6} \quad (S17)$$

where

$$B_{CuS} = \frac{18 d_{eff}^{CuS'^2} \omega^2 L_{CuS}^2}{n_3^{CuS} n_2^{CuS} n_1^{CuS} \epsilon_0 c^3} \quad (S18)$$

and

$$I_{CuS}^{direct}(3\omega) = \frac{27 d_{eff}^{CuS(3)^2} \omega^2 I(\omega)^3 L_{CuS}}{n_3^{CuS} (n_1^{CuS})^3 c^3} \quad (S19)$$

Here $d_{eff}^{CuS'}$ is the effective second-order susceptibility tensor for CuS films in SFG, $d_{eff}^{CuS(3)}$ is the effective third-order susceptibility tensor for CuS films in direct THG and $n_3^{CuS}$ is the refractive index of the CuS films at the third harmonic frequency.

Thus, the total THG intensity from the Au/CuS heterostructure films can be expressed as:

$$I_{Au/CuS}(3\omega) = [I_{pump}^{Au}(3\omega) + I_{pump}^{CuS}(3\omega)] + \frac{(A_{CuS-Au} B_{Au} + A_{Au-CuS} B_{CuS}) I(\omega)}{r^6} \quad (S20)$$

where

$$I_{pump}^{Au}(3\omega) = I_{Au}^{direct}(3\omega) + B_{Au}I(\omega)I_{pump}^{Au}(2\omega) \qquad (S21)$$

And

$$I_{pump}^{CuS}(3\omega) = I_{CuS}^{direct}(3\omega) + B_{CuS}I(\omega)I_{pump}^{CuS}(2\omega) \qquad (S22)$$

Clearly, the THG in the Au/CuS films consists of two parts: the first part is the incoherent sum of THG from Au and CuS nanoparticles without plasmonic coupling, including both direct THG and cascaded THG. The second term is the cascaded THG from Au and CuS nanoparticles, produced by the SFG between the coupled second-harmonic resonance and a fundamental pump photon. This route exhibits a characteristic inverse-sixth power dependence, just as SHG because it also depends linearly on the strength of the dipole-dipole interaction at 2ω.

## 2.3 FDTD Simulation of Plasmonic Coupling Between the Au and CuS Nanoparticles

In the FDTD simulations, we calculated the near-field enhancement when the Au and CuS nanoparticles are modeled in close proximity and excited by the 1050nm input light. According to equation (S6), the near-field strength exhibits an inverse-sixth power dependence. Therefore, the square of the near-field enhancement $\left(\frac{E_{nf}}{E_{pump}}\right)^2$ directly obtained from the simulation, provides the strong evidence for exploring the HIPRET process.

As a conclusion, in this section, we have developed a comprehensive theoretical model to explain the enhancement effect of the HIPRET on SHG and THG in the Au/CuS heterostructured films. Our model demonstrates that the plasmon-plasmon coupling between the second-harmonic of the CuS plasmon and the fundamental Au plasmon follows an inverse-sixth power distance dependence, consistent with the behavior of dipole-dipole interactions. Through SHG and THG experiments, we have observed that the energy transfer between CuS and Au plasmons significantly enhances harmonic generation. The energy transfer occurs via a bidirectional process, where both components in the heterostructure benefit from the resonant energy coupling. The FDTD simulation results provide further support for the inverse-sixth power law, confirming that the near-field enhancement in the Au/CuS system is a key factor in boosting harmonic generation efficiency.